\documentclass[
 reprint,
 amsmath,
 amssymb,
 aps,
]{revtex4-1}

\usepackage[pdftex]{graphicx}
\usepackage{wasysym}
\usepackage{amsfonts}
\usepackage{ifsym}
\usepackage[usenames,dvipsnames]{xcolor}
\usepackage{color}
\usepackage{graphicx}
\usepackage{dcolumn}
\usepackage{bm}
\usepackage{hyperref}

\newcommand{\beq}{\begin{equation}}
\newcommand{\eeq}{\end{equation}}
\newcommand{\beqy} {\begin{eqnarray}}
\newcommand{\eeqy} {\end{eqnarray}}

\def\({\left(}
\def\){\right)}
\def\[{\left[}
\def\]{\right]}
\def\<{\langle}
\def\>{\rangle}
\def\pd{\partial}
\def\ra{\rightarrow}

\def\a{\alpha}

\def\d{\delta}

\def\th{\theta}

\def\o{\omega}

\def\t{\tau}
\def\s{\sigma}

\def\ss{\boldsymbol\sigma}
\def\tt{\boldsymbol\tau}

\def\pd{\partial}
\def\pdb{\bar{\partial}}

\begin{document}

\title{
Smooth Wilson loops from the continuum limit of null polygons
}

\author{Jonathan C Toledo$^{\,1,2}$}

\vspace{15mm}

\affiliation{
\vspace{5mm}
$^{1}$Centro de F$\acute{\imath}$sica do Porto, Departamento de F$\acute{\imath}$sica e Astronomia \\
Faculdade de Ci$\hat{e}$ncias da Universidade do Porto, Rua do Campo Alegre 687, 4169-007 Porto, Portugal} 
\affiliation{
\vspace{5mm}
$^{2}$Perimeter Institute for Theoretical Physics \\
Waterloo, Ontario N2L 2Y5, Canada 
}

\begin{abstract}
\vspace{3mm}
We present integral equations for the area of minimal surfaces in $AdS_3$ ending on generic smooth boundary contours. The equations are derived from the continuum limit of the AMSV result for null polygonal boundary contours.  Remarkably these continuum equations admit exact solutions in some special cases.  In particular we describe a novel exact solution  which interpolates between the circle and 4-cusp solutions.
\end{abstract}

\maketitle

%%%%%%%%%%%%%%%%%%%%%%%%%%%%%%%%%%%
%%%%%%%%%%%%%%%%%%%%%%%%%%%%%%%%%%%
%%%%%%%%%%%%%%%%%%%%%%%%%%%%%%%%%%%
%%%%%%%%%%%%%%%%%%%%%%%%%%%%%%%%%%%

\section{Introduction} 
\vspace{-2mm}

The study of minimal surfaces goes back at least as far as the time of Lagrange who in 1768 considered the problem:  Find a surface of least area ending on a given closed contour \cite{Lagrange}.  This problem grew into an entire field of mathematics known as minimal surface theory and has occupied the attention of mathematicians and physicists alike for over two centuries (see \cite{review} for a recent review).  

Although historically most effort has focused on surfaces embedded in flat space, recent years have seen a shift in attention to minimal surfaces embedded in special curved spaces as a result of the so-called AdS/CFT correspondence \cite{Maldacena:1998re, Gubser:1998bc, Witten:1998qj}.  In particular, it was shown in the seminal work \cite{JMLoops98} that the expectation value of  Wilson Loops in certain strongly coupled gauge theories can be computed in terms of minimal surfaces in Anti-de Sitter (AdS) spacetime.  This provides an invaluable tool for the study of gauge theories as Wilson Loops are one of their most fundamental observables.  Moreover, the relation with minimal surfaces holds at large values of the coupling parameter where conventional perturbative techniques fall short.  

Although the mathematical statement of the problem is simple and perfectly well-posed -- compute the area of the minimal surface ending on a given closed contour at the boundary of AdS -- in practice this is a challenging task.  A hand full of exact solutions exist in cases where the boundary curve has an exceptional amount of symmetry.  For example, for closed loops in Euclidean AdS$_3$ one can construct solutions for a circular \cite{CircularLoop} and lens-shaped \cite{Lens} boundary curve.   There is also a beautiful method for constructing solutions parameterized by Riemann surfaces using theta-function techniques \cite{ThetaFunctions}.  In the case of closed spacelike loops in Minkowskian AdS the area can be computed exactly for the circle and the 4-cusp \cite{4cusp} solution, for example.

In a parallel development, recent years have witnessed a boom in our understanding of special types of surfaces in AdS based on the integrability of the underlying sigma model.  Thus far, integrability based techniques have successfully been applied to describe surfaces which approach the boundary at spikes -- relevant for the study of correlation functions \cite{CorrFunc} -- as well as surfaces which approach the boundary along generic null polygons -- relevant for the study of Wilson Loops and scattering amplitudes \cite{4cusp, WilsonLoop, AMSV}.    These results hold only in the strong coupling limit of the theory, where the problem becomes one of classical strings moving in AdS.

\begin{figure}[t]
\includegraphics[width=0.95\linewidth]{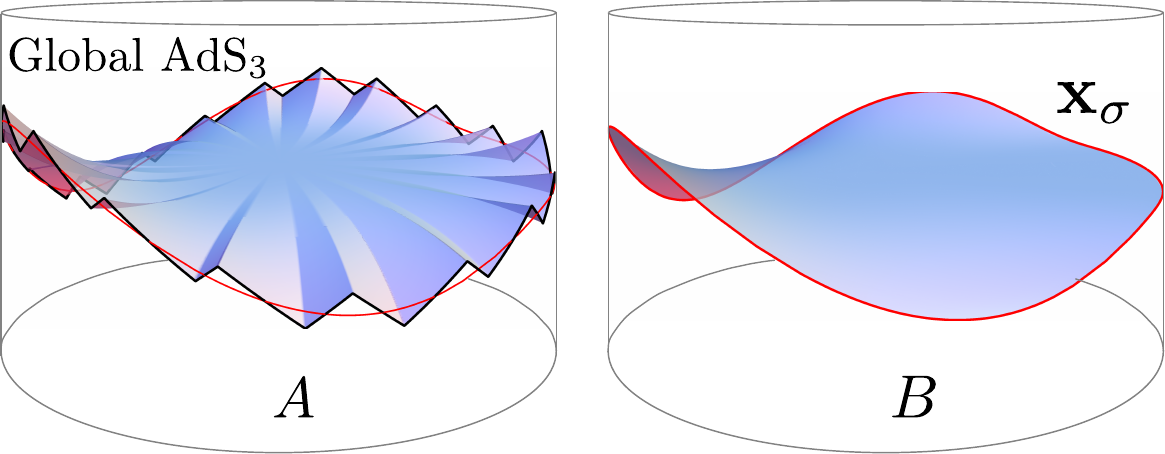}
\caption{\emph{Artistic depiction of minimal surfaces in global AdS}.  Surface $A$ ends on a null polygon at the boundary of global AdS$_3$, which is indicated by the gray cylinder.  As the number of cusps becomes large surface $A$ limits to surface $B$, which ends on the smooth curve $\bf{x}_{\s}$ shown in red.}\label{global_AdS_plt}
\end{figure} 

Two features of the integrability-based approach should be emphasized.   First, this approach is very economical in that one directly computes the minimal area without ever needing to know the shape of the embedding surface.  This gives an enormous analytical and numerical advantage in the treatment of the problem.  Second, and perhaps most important, they provide a manifestly integrable formulation of the purely \emph{classical} worldsheet problem.  In the past this was the key step in determining the solution to the full \emph{quantum} problem.  In particular, this was the case in \cite{KMMZ} and \cite{AFS} for the spectral problem and \cite{AMSV} and \cite{OPE} for the all-loop description of scattering amplitudes.  

Inspired by these unprecedented achievements, we set our sights on an all-loop description of smooth Wilson Loops in $\mathcal{N}=4$ Super Yang Mills Theory.  The first step toward this end is to develop a manifestly integrable formulation of the classical worldsheet problem that emerges at strong coupling, which is the aim of this paper.   In this work we initiate a systematic integrability-based study of minimal surfaces in AdS which end on smooth curves at the boundary.   The main observation is the simple and well-known fact that any smooth curve can be approximated to arbitrary accuracy by a sequence of null segments (see figure \ref{global_AdS_plt}).   Thus we can start with the results of \cite{AMSV} for null polygons and compute the minimal area of \emph{any} smooth (simply connected) boundary curve by performing a careful continuum limit.  The result of this continuum limit, and the main result of this paper, is a novel set of integral equations whose solution yields the area of minimal surfaces ending on smooth curves at the boundary of AdS.  We will refer to these equations as the Continuum Thermodynamic Bethe Ansatz  equations or simply CTBA equations since they are the continuum analog of TBA equations derived in \cite{AMSV} for null polygons.  

We describe our method for computing the area of minimal surfaces in section \ref{eqns}.  In section \ref{checks} we study a special exact solution of the CTBA and in doing so we introduce a new AdS string solution whose area can be computed exactly.  We do \emph{not} provide a derivation of the results herein which will be given elsewhere \cite{in_prep_1}.  We stress however, that the derivation is not simply a matter of taking a large number of cusps in the results of \cite{AMSV} as figure \ref{global_AdS_plt} may suggest.  It turns out that the limit is rather subtle.  One must perform an elaborate analytic continuation of the equations in \cite{AMSV} before taking the number of cusps to be large.  Only then does one describe generic smooth curves in this limit.  This analytic continuation is a beautiful mathematical problem with many connections to the topic of wall-crossing \cite{GMN} as we will describe in \cite{in_prep_1}.   In the appendix we introduce a numerical implementation of our method for computing minimal areas.  This allows us to demonstrate that the CTBA is not only a powerful tool for analytics, but is also a useful computational tool.  It also affords the opportunity to perform a final check of the equations presented here.  In the appendix we show that results obtained from the CTBA agree with those obtained from direct numerical integration of the string equations of motion.  %:)%
%%%%%%%%%%%%%%%%%%%%%%%%%%%%%%%%%%%
%%%%%%%%%%%%%%%%%%%%%%%%%%%%%%%%%%%
%%%%%%%%%%%%%%%%%%%%%%%%%%%%%%%%%%%
%%%%%%%%%%%%%%%%%%%%%%%%%%%%%%%%%%%
\section{Minimal area for smooth curves}\label{eqns}
\begin{figure}[t]
\centering
\includegraphics[width=0.65\linewidth]{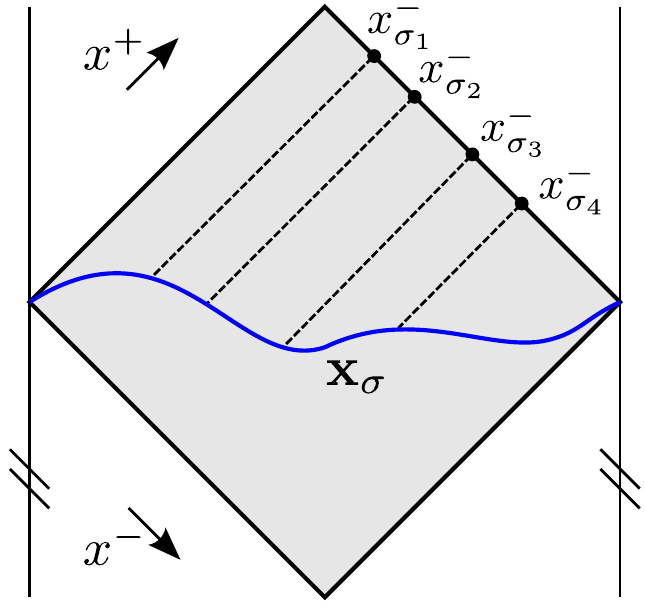}
\caption{\emph{Boundary curve} $\bf{x}_{\s}$ \emph{and cross ratios.}  To go back to global AdS one simply imagines wrapping the Poincar\'e patch (in gray) on the cylinder.  We indicate the lightcone directions $x^{\pm}$ and show four points along the $x^{-}$ direction.  From four such points one can form the cross ratios $X^-_{\s_1\s_2\s_3\s_4}$. The regularized area \eqref{Afull} depends only on the conformal invariant quantities $X^{\pm}_{\s_1\s_2\s_3\s_4}$.  Here we will exclusively work in terms of  bi-local cross ratios formed by taking the limit $\s_2 \ra \s_1$, $\s_4 \ra \s_3$ in these full 4-point cross ratios. }\label{XsigmaPoincarePatch}
\end{figure}

\vspace{-2mm}
In this section we present our method for computing minimal areas using integrability.  We are interested in surfaces of minimal area which end on a smooth curve $\bf{x}_{\s}$ at the $R^{1,1}$ boundary of AdS$_3$ as shown in figure \ref{XsigmaPoincarePatch}.  We consider everywhere spacelike curves which wrap the AdS cylinder once.  In this case the worldsheet is euclidean and has the topology of a disk.  We describe the worldsheet with the usual complex coordinates  $(z,\bar{z})$.  In the $z$-plane there is a curve $z_{\s}$ which maps to the spacetime boundary curve $\bf{x}_{\s}$ and the region inside of this curve maps onto the worldsheet (see figure \ref{MassesAndIntMat}).  The curve $z_{\s}$, although unphysical, will be very important in our construction as it will parameterize the physical boundary curve $\bf{x}_{\s}$.\\
\indent The full area has a well understood arc-length divergence and we construct the usual regularized area defined by
\beq\label{Afull}
A=A_{\text{full}}-\frac{L}{\mathcal{E}}
\eeq
where $L$ is the arc-length of the boundary curve and $\mathcal{E}$ is the standard cut-off in the AdS radial direction.  The regularized area $A$ is a conformal invariant quantity 
and thus does not depend on the boundary curve itself, but rather on the conformal invariant data encoded in this curve.  We construct  these conformal invariants from points along the boundary curve by forming cross ratios as shown in figure \ref{XsigmaPoincarePatch}.  The full set of cross ratios is a hugely over-complete basis.  A nicer basis (although still over-complete) is the more local set of cross ratios obtained by colliding pairs of points as described in figure \ref{XsigmaPoincarePatch}.  We will work exclusively in terms of these bi-local cross ratios which can be thought of as the conformal invariant notion of a distance. 

The main ingredient of the integrability based approach is to deform these cross ratios by the introduction of a spectral parameter $\th$.  We denote the deformed bi-local cross ratios as $\hat{y}_{\ss}(\th)$ where the boldface index stands for the pair of indices $\ss=(\s_1,\s_2)$.  This deformation is useful because, remarkably, the area \eqref{Afull} can be computed directly in terms of these y-functions, which can in turn be computed by careful analysis of their $\th$-analyticity properties!  In this way the area can be computed without ever knowing the actual shape of the minimal surface, as mentioned above.  The y-functions can be written in terms of a deformed boundary curve $x_{\s}^{\th}$ as
\beq\label{y functions}
\hat{y}_{\ss}(\th) = \frac{\pd x_{\s_1}^{\th}\pd x_{\s_2}^{\th}}{(x_{\s_1}^{\th}-x_{\s_2}^{\th})^2}
\eeq
The scalar function $x_{\s}^{\th}$ conveniently encodes both the $x^{+}$ and $x^{-}$ components of the boundary curve as $x^{\th=0}_{\s}=x^+_{\s}$ and $x^{\th=i\pi/2}_{\s}=x^-_{\s}$.  Thus we recover the physical cross-ratios as
\beq\label{Xratios}
\hat{y}_{\ss}(0)=\frac{\pd x^{+}_{\s_1}\pd x^{+}_{\s_2}}{(x^{+}_{\s_1}-x^{+}_{\s_2})^2},\;\;\;\;\; \hat{y}_{\ss}(i\pi/2)=\frac{\pd x^{-}_{\s_1}\pd x^{-}_{\s_2}}{(x^{-}_{\s_1}-x^{-}_{\s_2})^2}
\eeq
The function $x^{\th}_{\s}$ is only defined up to global conformal transformations, which can be $\th$-dependent.  As such, its analytic properties are far less constrained than those of the y-function \eqref{y functions} which is independent of such transformations.  The cross ratios \eqref{y functions}-\eqref{Xratios} also have nice properties under $\s$-reparameterization, transforming with weight two.  

Our main goal in this section will be to write the equations which determine the $y$-functions \eqref{y functions} as well as the functional which computes the area from them.
%%%%%%%%%%%%%%%%%%%%%%%%%%%%%%%%%%%
%%%%%%%%%%%%%%%%%%%%%%%%%%%%%%%%%%%
\subsection{Continuum Thermodynamic Bethe Ansatz}\label{sec_CTBA}
\begin{figure}[t]
\includegraphics[width=0.65\linewidth]{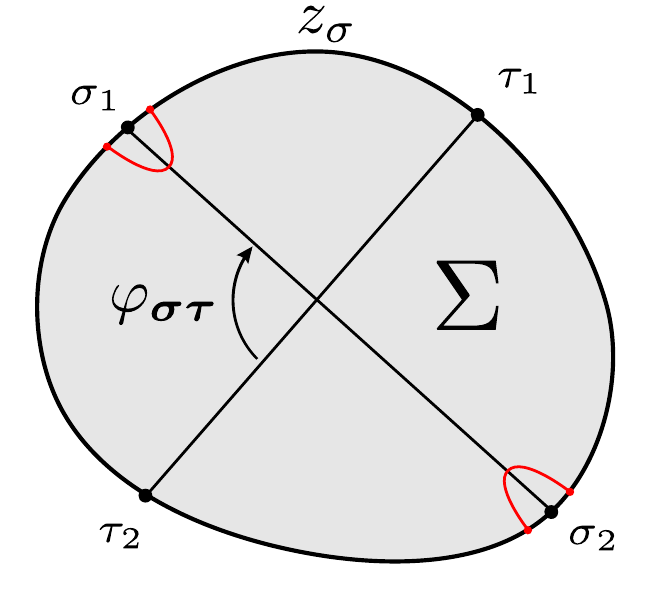}
\caption{\emph{Worldsheet curve $z_{\s}$ and action of $K$.}  The interior region $\Sigma$ maps to the embedding surface in the interior of AdS and $z_{\s}$ maps to $\bf{x}_{\s}$ under the string embedding.  This sort of diagram also serves as a useful pictorial representation of the kernel \eqref{CTBA K}.  We represent a y-function $y_{\ss}$ with a straight line connecting $z_{\s_1}$ and $z_{\s_2}$ and similarly for $y_{\tt}$.  The $\tt$-integration in \eqref{CTBA K} is only over $\tt$ such that there is an intersection of these two lines.  The phase $\varphi_{\ss\tt}$ in \eqref{CTBA K} is the angle between these two lines.  We schematically indicate the ``pinch" region of the integration by the curved red lines.  From \eqref{y functions} we see that $y_{\tt}$ develops a singularity $(\t_2-\t_1)^{-2}$ in this region.  
 }\label{MassesAndIntMat}
\end{figure}

Let us now describe how we fix the $y$-functions \eqref{y functions} and how we compute the area \eqref{Afull} in terms of them.  The solution is parametric.  We parameterize the problem in terms of the complex curve $z_{\s}$ which maps to the boundary curve $\bf{x}_{\s}$ under the string embedding.    The $y$-functions are fixed by this data according to the equations
\beq\label{CTBA}
y_{\boldsymbol\sigma}(\theta)=y_{\boldsymbol\sigma}^{\text{circ}}e^{-4|z_{\boldsymbol\sigma}|\cosh \theta - K_{\boldsymbol\sigma}^{\boldsymbol\tau} \star (y_{\boldsymbol\tau}-y_{\boldsymbol\tau}^{\text{circ}}) }
\eeq
where we use the notation $z_{\ss}=z_{\s_2}-z_{\s_1}$ and the action of the kernel  is given by 
\beq\label{CTBA K}
K_{\ss}^{\tt } \star f_{\tt}=\!\!\! \int\limits_{\s_2}^{\s_1+2\pi}  \!\!\! d\tau_2  \int\limits\limits_{\s_1}^{\s_2}  \! d\tau_1\! \int\limits_{\!\mathbb{R}} \! \frac{d \theta'}{i\pi} \frac{f_{\tt}(\theta')}{ \sinh (\theta\!-\!\theta'\!+\! i \varphi_{\ss}\!\!-\! i \varphi_{\tt})} 
\eeq
The unhatted $y$-functions are related to the hatted ones via $y_{\ss}(\theta)=\hat{y}_{\ss}(\theta+i\varphi_{\ss})$ where $\varphi_{\ss}=\text{arg} (z_{\ss})$.  The $\tt$ integration has the pictorial representation shown in figure \ref{MassesAndIntMat}.   The role of $y^{\text{circ}}$ is to regulate the short distance singularity that occurs in the ``pinching" region of the $\tt$-integration where $\t_1$ and $\t_2$ both approach $\s_1$ or $\s_2$ (see figure \ref{MassesAndIntMat}). This $\th$-independent regulating function is not unique, and the solution of \eqref{CTBA} does not depend on the choice of $y^{\text{circ}}$.  We can write it as
\beq\label{y circ}
y^{\text{circ}}_{\ss} \equiv \frac{\pd x_{\s_1}^{\circ}\pd x_{\s_2}^{\circ}}{(x_{\s_1}^{\circ}-x_{\s_2}^{\circ})^2}
\eeq 
where $x^{\circ}_{\s}$ is any smooth, monotonic curve running from $-\infty$ to $+\infty$.  What is important is that because of its bilocal form \eqref{y circ} automatically has the proper short-distance singularity in $\s$ such that the integration in \eqref{CTBA K} is regulated in the pinch region.   Additionally, it has the correct transformation properties such that \eqref{CTBA} is parameterization invariant.   Alternatively, one could write \eqref{CTBA}-\eqref{CTBA K} without using such a regularizing function by introducing some infinitesimal regulating $\epsilon$ factors.  

We call \eqref{CTBA} the Continuum Thermodynamic Bethe Ansatz  (CTBA) equations.  They are the continuum analog of the TBA equations derived in \cite{AMSV} for null polygons.  
There are limits and special cases where \eqref{CTBA} can be solved exactly and we shall provide some examples below.  However, for general smooth curves one will need to integrate the CTBA numerically.   We develop an efficient numerical implementation of \eqref{CTBA} in the appendix. A demonstration of these numerics is given in figure \ref{AreaIts}.
%%%%%%%%%%%%%%%%%%%%%%%%%%%%%%%%%%%
%%%%%%%%%%%%%%%%%%%%%%%%%%%%%%%%%%%
\subsection{Area and kink solution}\label{sec_Area}
Once the functions $y_{\ss}(\theta)$ are obtained from \eqref{CTBA} we compute the area as
\beq\label{Area1}
A=-2\pi - 4 A_{\Sigma} - A_{\text{free}}
\eeq
Here $A_{\Sigma}$ is the area enclosed by $z_{\s}$ and $A_{\text{free}}$ is given by 
\beq\label{Afree}
A_{\text{free}} = \frac{1}{\pi} \int\limits_{0}^{2\pi} d\sigma_1 \int\limits_{0}^{2\pi} d\sigma_2 \! \int\limits_{\mathbb{R}} d\theta \,|z_{\boldsymbol\sigma}|e^{-\theta}(y_{\boldsymbol\sigma}-{y}^{\text{kink}}_{\boldsymbol\sigma})
\eeq
where we have defined
\beq\label{kink}
y^\text{kink}_{\boldsymbol\sigma}(\th)
= \pd z_{\s_1}\pd z_{\s_2} e^{-2(\theta+i \varphi_{\ss})}\mbox{csch}(e^{-\theta}|z_{\boldsymbol\sigma}|)^2
\eeq
The appearance of $y^{\text{kink}}$ in \eqref{Afree} is a consequence of the regularization of the area.   Roughly speaking, in \eqref{Afree} we regulate the divergence of the full area \eqref{Afull} of a minimal surface with boundary curve arc-length $L$ by subtracting off the area of the solution ending on a circle of circumference $L$.   As we will explain momentarily, this subtraction is implemented precisely by the subtraction of $y^{\text{kink}}$ in \eqref{Afree}.   Since we are subtracting off the area of the solution ending on a circle, we must add back the finite part of this area to obtain the correct prescription \eqref{Afull}.  This explains the appearance of the $-2\pi$ in \eqref{Area1}  which is precisely the finite part of the area of the minimal surface ending on a circle \cite{CircularLoop}.  

In fact, it is a useful exercise to recover from \eqref{Afree} the famous result  \cite{CircularLoop} for the area of the circular wilson loop.  This also allows one to clearly see the physical meaning of $y^{\text{kink}}$. The circular loop corresponds to the case in which the world-sheet curve collapses to a point and thus the distances $z_{\ss}$ vanish in the sources of the CTBA.   This is the analog of the high-temperature limit in usual TBA nomenclature \cite{highT} and the treatment of this limit is somewhat analogous.  The $y$-functions form broad plateaus of width  $\theta \sim \log |z_{\ss}|$ and at the edge of the plateau there is a ``kink" where the $y$-functions decay rapidly to zero.   Because the integrand of $A_{\text{free}}$ is proportional to $|z_{\ss}|e^{-\th}$ the only possible non-vanishing contribution comes from this kink region.   Thus, we study this limit by shifting the $\theta$ variable $\th \ra \th+\log |z_{\ss}|$ in the CTBA to focus on the kink.   The resulting kink-TBA equations are exactly the same as the original TBA  equations but with the source modified as $\cosh \th \ra 1/2 e^{-\th}$.  We claim that \eqref{kink} is the exact solution of the CTBA \eqref{CTBA} after this replacement has been made in the source.   Because the exact behavior of the $y$-function in this kink region is given by $y^{\text{kink}}_{\ss}$ in \eqref{kink} it turns out that $A_{\text{free}}$ vanishes completely as the boundary curve $\bf{x}_{\s}$ approaches a circle.  The contribution $A_{\Sigma}$ vanishes by definition (recall that it is simply the area enclosed by $z_{\s}$) and thus from \eqref{Area1} we recover the expected result for the circular wilson loop \cite{CircularLoop}
\beq\label{circle area}
A^{\text{circ}} = -2\pi.
\eeq

Finally, we note that the existence of the exact solution \eqref{kink} of the CTBA in the limit of a circular boundary curve is quite surprising.  The analogous limit in usual TBA equations (what is usually called the high-Temperature or conformal limit \cite{highT}) one can only find the height and approximate width of the plateau region, but there seems to be no hope of finding the exact shape of the kink which is crucial for the high-T expansion \cite{SLYM}.   We find it amazing that this nonlinear (triple) integral equation has an exact solution for generic $z_{\s}$, especially one so simple as \eqref{kink}.  What is even more remarkable is that one can find exact solutions to the \emph{full} CTBA equations (i.e. not just in the high-T limit) as will discuss in the following section.  The existence of such exact solutions is a novel feature of the continuum TBA as there are presently no exact solutions of any usual TBA equation, even for the simplest models such as the Scaling Lee-Yang Model \cite{SLYM}.  \\
\indent Equations \eqref{CTBA}-\eqref{Afree} are the main result of this paper.  They provide an integrability based method of computing minimal areas in $AdS_3$ for rather generic boundary curves.  In the following section we study an exact solution of these equations.  We give a numerical implementation in the appendix.
%%%%%%%%%%%%%%%%%%%%%%%%%%%%%%%%%%%
%%%%%%%%%%%%%%%%%%%%%%%%%%%%%%%%%%%
%%%%%%%%%%%%%%%%%%%%%%%%%%%%%%%%%%%
%%%%%%%%%%%%%%%%%%%%%%%%%%%%%%%%%%%
\section{An exact solution}\label{checks}
\vspace{-2mm}
\begin{figure}[t]
\includegraphics[width=0.75\linewidth]{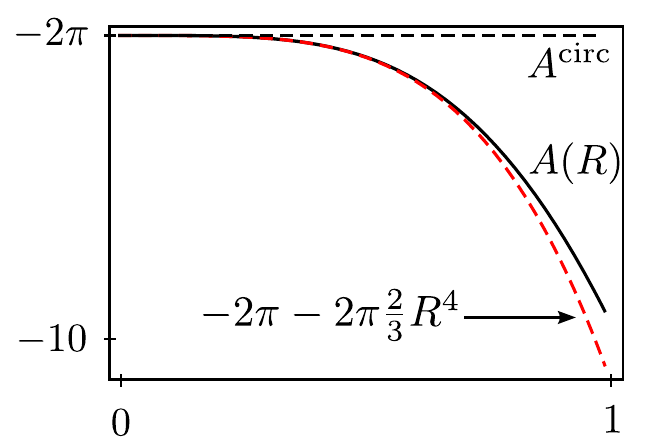}
\caption{Plot of $A(R)$ given in \eqref{MathieuArea}.  The solid black curve is $A(R)$.  The horizontal dashed curve shows the value $A^{\text{circ}}=-2\pi$.  Note that $A(R)\le A^{\text{circ}}$ with the equality holding only at $R=0$.  The dashed red curve is the small $R$ expansion of $A(R)$.  One can check that the small $R$ expansion is in precise agreement with the wavy line approximation \cite{Wavy}.  The large $R$ expansion (not shown here) is also consistent with the approach to the 4 cusp solution as it develops divergences in $R$ which one can interpret as cusp divergences.}\label{AreaR}
\end{figure}
\begin{figure}[t]
\centering
\includegraphics[width=0.75\linewidth]{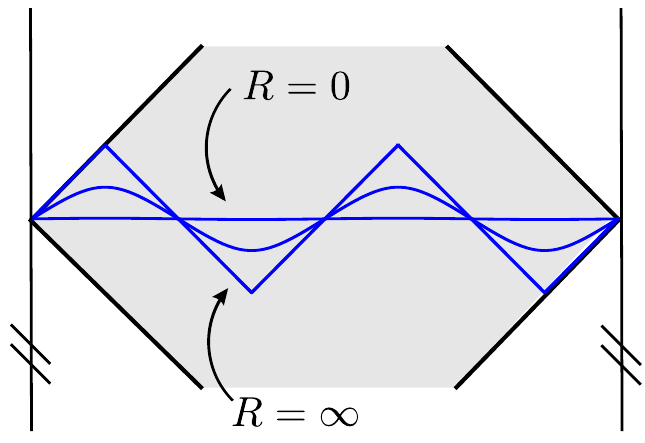}
\caption{\emph{Mathieu solution}.  Boundary curves for $R\!=\!0, 3/4, \infty$ in the Poincar\'e Patch.   For this plot we use the convenient conformal frame $x_{\s}^{+} = x_{\s}^{\th=0}$, $x_{\s}^{-}= (-1+x_{\s}^{\th=i\pi/2})/(1+x_{\s}^{\th=i\pi/2})$ which differs from the one in the main text by a simple conformal transformation.}\label{MathieuInPP}
\end{figure}

In this section we study a special 1-parameter family of exact solutions of \eqref{CTBA} which interpolate between the surface ending on a circular loop and the surface ending on a null-square or what is also known as the ``4-cusp solution".  These solutions correspond to the family of worldsheet curves $z_{\s}=R \, e^{i \sigma}$.  Quite surprisingly the CTBA \eqref{CTBA} can be solved exactly in this case and the $y$-function is given by \eqref{y functions} with 
\beq\label{MathieuSol}
x_{\s}^{\theta} = \frac{M\!c(i \th + \s)}{M\!s(i \th + \s)}
\eeq
where $M\!c$ and $M\!s$ are Mathieu cos and sin functions \cite{FootNote_MathieuImp}.  From the exact $y$-function we compute the regularized area using \eqref{Afree}-\eqref{kink}.  The result is
\beq\label{MathieuArea}
A(R) = -2 \pi - 2\pi\(1/4-a(R)\)
\eeq    
where $a(R)$ is the Mathieu Characteristic \cite{FootNote_MathieuImp}.  We plot the area $A(R)$ in figure \ref{AreaR}.   Although $z_{\s}$ has rotational symmetry, the spacetime surface does not (except at $R=0$ which corresponds to the circle as discussed in section \ref{eqns}).  Indeed, as $R$ tends to infinity the boundary curve approaches a null square.   For intermediate values of $R$ the boundary curve is some non-trivial closed curve on the cylinder which interpolates between these limiting cases as shown in figure \ref{MathieuInPP}.  The fact that the boundary curve $\bf{x}_{\s}$ approaches the circle and null square in the limits  $R\ra0$ and $R\ra \infty$ respectively is not a special feature of the circular worldsheet curve.  Indeed, this feature is universal.  For any curve $z_{\s}$ the boundary curve approaches the circle or null square as $z_{\s}$ collapses to a point or is blown to infinity respectively. 

We emphasize again that the existence of exact solutions of \eqref{CTBA} is a remarkable and novel feature of the continuum limit.  There is no analog for usual TBA equations for which a single non-trivial exact solution is yet to be discovered (e.g. in the case of null polygons).  

Finally, let us now comment on how this solution was obtained. Underlying the integrability construction described above is the linear problem associated with the string equations of motion in AdS$_3$.  It turns out that for a circular worldsheet curve $z_{\s}$, this linear problem is equivalent to one that was recently studied \cite{Luky_YY_2011} in the context of the Sinh-Gordon model.  Remarkably, in \cite{Luky_YY_2011} the exact wronskian ($Q$-function) of that linear problem was constructed.  Starting from this $Q$-function it is possible to construct the solution \eqref{MathieuSol}.  Here we only compute the area and the boundary contour, however perhaps it is possible to actually construct the full embedding surface.

This concludes the presentation of analytical results and we will now discuss future directions and conclude.  
%%%%%%%%%%%%%%%%%%%%%%%%%%%%%%%%%%%
%%%%%%%%%%%%%%%%%%%%%%%%%%%%%%%%%%%
%%%%%%%%%%%%%%%%%%%%%%%%%%%%%%%%%%%
%%%%%%%%%%%%%%%%%%%%%%%%%%%%%%%%%%%
\section{Discussion}\label{discussion}
\begin{figure}[t]
\centering
\includegraphics[width=0.85\linewidth]{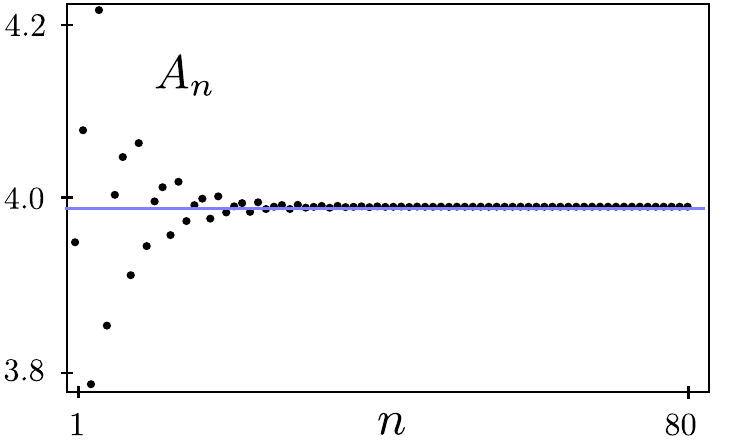}
\caption{ \emph{Convergence of CTBA.}  Area $A_{n}$ computed from \eqref{Area1}-\eqref{Afree} from the $n$th iteration of the CTBA starting from the initial iterate $y^{\text{reg}}_{(0)} = 0$.  We use the boundary curve $z_{\s} = e^{i \s} + 1/5 \, e^{2 i \s}$.   The details of the numerical integration are given in the appendix \ref{CTBA Num}.  For the parameters used to produce this plot, see footnote \cite{numerics_parameters}.    The CTBA converges to $A^{\text{CTBA}} = 3.990...$.  The blue line is at $A^{\text{EOM}}=3.989...$ and indicates the value obtained by direct numerical integration of the Pohlmeyer-reduced string equations of motion which we explain in appendix \ref{Pohl Num}. }\label{AreaIts}
\end{figure}

\vspace{-2mm}
In this work we develop an integrability-based method for computing the area of minimal surfaces in AdS which end on smooth curves at the boundary.   Our main result is the set of integral equations \eqref{CTBA}-\eqref{Afree}.  These integral equations, which we dubbed the CTBA,  provide a powerful tool for analytic study of minimal surfaces.  The CTBA also provides a powerful tool for numerics.  In the appendix we develop an algorithm for numerically integrating these equations.  It efficiently reproduces the results obtained from a brute force numerical integration of the string equations of motion as shown in figure \ref{AreaIts}.  

Perhaps the most exciting aspect of this work is the myriad possibilities for future directions.  First, it would be interesting to generalize the results here to minimal surfaces in the full AdS$_5$.  It should be possible to do this by following the same steps used to derive the equations presented here, but starting from the AdS$_5$ version of the AMSV equations  \cite{AMSV}.  We will comment more on this when we present the derivation of the CTBA \cite{in_prep_1}.  Another interesting direction would be to adapt the results described above for the case of \emph{Euclidean} AdS.  This would be of interest in the study of entaglement entropy and should also reveal a fascinating connection between (C)TBA and theta functions \cite{ThetaFunctions}.   We have made much progress towards this end and will present the results elsewhere.

First and foremost, we see the results presented here as the first step in the study of smooth Wilson loops in $\mathcal{N}=4$ at any value of the t'Hooft  coupling.  Although a daunting task, history has taught us that it is indeed possible as demonstrated by the exact solution of the spectrum problem as well as recent results from the OPE of null polygonal Wilson loops.  In both cases, the first step was to identify integrability in the extreme strong and weak coupling limits.  In this paper we achieve the former.  A natural next step is to study the weak coupling problem where there have been remarkable advances in the study of null polygonal Wilson loops.  It would be very interesting to study the continuum limit of these perturbative results and to try to identify some hidden integrability structure present at both weak and strong coupling. 
 
\noindent

{\it Acknowledgements:} It is a pleasure to thank Pedro Vieira and Amit Sever for pointing me to this problem and for an extended period of collaboration as well as many valuable discussions.  I would also like to thank J. Caetano, N. Doroud, S. Komatsu, D. Gaiotto, N. Gromov, F. Alday, M. Kruczenski, B. Basso and J. Penedones for useful discussions.
This work was supported in part by the [European Union] Seventh Framework Programme [FP7-People-2010-IRSES] under grant agreements No 269217, 317089.  \emph{Centro de Fisica do Porto} is partially funded by the Foundation for Science and Technology of Portugal (FCT).
%%%%%%%%%%%%%%%%%%%%%%%%%%%%%%%%%%%
%%%%%%%%%%%%%%%%%%%%%%%%%%%%%%%%%%%
%%%%%%%%%%%%%%%%%%%%%%%%%%%%%%%%%%%
%%%%%%%%%%%%%%%%%%%%%%%%%%%%%%%%%%%
\begin{appendix}
%%%%%%%%%%%%%%%%%%%%%%%%%%%%%%%%%%%
%%%%%%%%%%%%%%%%%%%%%%%%%%%%%%%%%%%
%%%%%%%%%%%%%%%%%%%%%%%%%%%%%%%%%%%
%%%%%%%%%%%%%%%%%%%%%%%%%%%%%%%%%%%
\section{Numerics}\label{numerics}
\vspace{-2mm}
In this appendix we turn to a numerical study of the CTBA \eqref{CTBA}.  The main purpose of this appendix is develop the proper numerical techniques to integrate the CTBA.  This allows us to demonstrate that these equations are more than a formal curiosity, but are also a practical tool for computing minimal areas.  The second goal of this section is to perform a final check of the CTBA as well as the numerical recipe that we present below.  For this we directly numerically integrating the (Pohlmeyer reduced) string equations of motion.  The results obtained from the two approaches agree well within the expected numerical error from each side.
%%%%%%%%%%%%%%%%%%%%%%%%%%%%%%%%%%%
%%%%%%%%%%%%%%%%%%%%%%%%%%%%%%%%%%%  
\subsection{CTBA}\label{CTBA Num}
First we must find a suitably regularized function for which to solve.  
There are two types of singularities in the region $\s_2 \sim \s_1$ which make $y_{\ss}(\th)$ an unsuitable function to use for numerics.  In the limit $\s_2 \ra \s_1$ with fixed $\th$ the y-function has the expansion
\beq\label{smalls}
y_{\ss}\sim \frac{1}{(\s_1-\s_2)^2}-\frac{1}{3}|\pd_{\s_1} z_{\s_1}|^2 e^{-2\th}-\frac{1}{3}|\pd_{\s_1} z_{\s_1}|^2 e^{+2\th} + ...
\eeq
where the $+... $ represents terms finite as $\s_2 \ra \s_1$ and $\th \ra \pm \infty$.  The first term presents one type of singularity which is simple to treat as it has only to do with short distances in $\s$.  One can remove it by forming combinations like $y_{\ss}-y^{\text{circ}}_{\ss}$.  The second, more difficult, type of singularity is due to the $\th$-dependent terms in \eqref{smalls} which reflect a subtle order of limits that occurs at small separation in $\s$ and large values of $\th$.  From \eqref{smalls} we see that if we take $\s_2 \ra \s_1$ and then take $\th$ large then the y-function diverges exponentially in $\th$.  On the other hand, if we first send $\th \ra \pm \infty$ and then take $\s_2 \ra \s_1$ in the y-function, the result will be zero.  To see this, first note that the expansion \eqref{smalls} is not valid in this limit.  For finite separation in $\s$ the dominant large $\th$ behavior is given by dropping the kernel term in \eqref{CTBA}.   From this it is clear that the y-function will go to zero double exponentially at large theta for any nonzero separation in $\s$.  In other words, there is a sort of ``boundary layer" at $\s_2-\s_1=0$ whose height diverges exponentially in $\th$. This divergent boundary layer is quite toxic for the numerics, but fortunately it can be regulated easily.  The key point is that the kink y-function captures this behavior exactly.  Thus a fully regulated function is given by
\beq\label{yreg}
y^{\text{reg}} = y - y^{\text{kink}}- y^{\text{akink}}+y^{\text{circ}}
\eeq
where $y^{\text{circ}} $ is given in \eqref{y circ} and $y^{\text{akink}}$ is given by
\beq
y^{\text{akink}}(\th)= y^{\text{kink}}(-\th)^{*}
\eeq
Note that $y^{\text{kink}}$ regulates the boundary layer at $\th \ra -\infty$, $y^{\text{akink}}$ regulates the boundary layer at $\th \ra +\infty$ and the double poles \eqref{smalls} at $\s_2\sim\s_1$ cancel between the terms on the right hand side of \eqref{yreg}.  Thus $y^{\text{reg}}$ is a good function for numerics.

The integral equation obeyed by $y^{\text{reg}}$ can be obtained by recalling that $y^{\text{kink}}$ obeys the CTBA \eqref{CTBA} but with $\cosh \th \ra 1/2 e^{-\th}$ (see the discussion after equation \eqref{kink}).  Similarly, $y^{\text{akink}}$ obeys the CTBA but with $\cosh \th \ra 1/2 e^{+\th}$.  Finally, $y^{\text{circ}}$ obeys the CTBA with $|z_{\ss}| \ra 0$ in the source.  Putting all of this together yields the $y^{\text{reg}}$ equation.
\beq\label{RegCTBA}
y^{\text{reg}}= - y^{\text{kink}}- y^{\text{\text{akink}}}+y^{\text{circ}} + \frac{ y^{\text{kink}}y^{\text{\text{akink}}}}{y^{\text{circ}} }e^{-K\star y^{\text{reg}}}
\eeq

Now that we have the equation \eqref{RegCTBA} suitable for numerics, we will describe our numerical method which is based on the usual iteration scheme used for TBA equations.  That is, we start with some initial guess for the y-function ($y^{\text{reg}}_{(0)}=0$, for example), plug it into the RHS of the CTBA, and integrate to produce an updated y-function.  We then repeat the process until it converges.  At first sight this seems painfully slow.  At each iteration one must perform the triple integration operation $K \star$ (see equation \eqref{CTBA K}) for each point in a suitable $\{\theta, \s_1,\s_2\}$ grid.  This grid typically contains around $10^4$ points and direct numerical implementation of $K\star$ (using Mathematica's NIntegrate, for example) takes several seconds for each point in this grid.  The end result is that each iteration takes a few hours.  Given that a few hundred iterations are needed for good convergence, this approach is clearly unreasonably slow.  This difficulty can be circumvented with the use of Fourier methods which allow one to convert $K \star$ into matrix multiplication such that each iteration (i.e. evaluation of the entire $\{\theta, \s_1,\s_2\}$ grid) can be performed in under a second even in an un-parallelized code.  We will now explain this in slightly greater detail.

First consider the $\th'$-integration.  Using convolution theorem we can write \eqref{CTBA K} as
\beq\label{conv}
\int\limits_{\s_2}^{\s_1+2\pi}  \!\!\! d\tau_2  \int\limits\limits_{\s_1}^{\s_2}  \! d\tau_1 \;  \mathcal{F}^{-1}_{\th}\!\! \[e^{\d_{\ss\tt} \, \o}  \mathcal{F}_{\o} \[(\pi \cosh \th')^{-1}\]\mathcal{F}_{\o}\[y^{\text{reg}}_{\tt}(\theta')\]\]
\eeq
where $\mathcal{F}$ and $\mathcal{F}^{-1}$ are forward and reverse Fourier transforms, $\o$ is the Fourier variable conjugate to $\th$, and \cite{ArgBranches}
\beq
\d_{\ss\tt}= \(\pi/2+\varphi_{\ss} - \varphi_{\tt}\)
\eeq
Now consider the $\t$ integration. The $\t$-variables are compact and thus we can expand $y^{\text{reg}}_{\tt}(\th')$ in modes as
\beq
\sum_{a_1 =-\infty}^{\infty}\sum_{a_2=0}^{\infty} e^{i \pi a_1 \a_1} \cos(\pi a_2 \a_2 )\, u_{a_1a_2}(\th')
\eeq
where we have introduced the useful shorthand
\beq
\a_1 = \frac{-2\pi+\t_2+\t_1}{2\pi}, \;\;\;\;\; \a_2 = \frac{\t_2-\t_1}{2\pi}
\eeq 
and made use of the $\t_2 \leftrightarrow \t_1$ symmetry of $y^{\text{reg}}$.  Plugging this expansion into \eqref{conv} gives
\beq\label{FTK}
K \star y^{\text{reg}} = \mathcal{F}^{-1}_{\th}\[\mathcal{C}_{\s_1\s_2}^{a_1 a_2}(\o) \mathcal{F}_{\o}\[(\pi \cosh \th')^{-1}\]  \mathcal{F}_{\o}\[u_{a_1a_2}(\th')\]\]
\eeq
where 
\beq
\mathcal{C}_{\s_1\s_2}^{a_1 a_2}(\o) = \!\!\! \int\limits_{\s_2}^{\s_1+2\pi}  \!\!\! d\tau_2  \int\limits\limits_{\s_1}^{\s_2}  \! d\tau_1  \, e^{i \pi a_1 \a_1} \cos(\pi a_2 \a_2) e^{\d_{\ss\tt} \o}
\eeq
The mode transfer matrix $\mathcal{C}$ is a fixed object:  it is computed once and for all for a given $z_{\s}$ and then is an input into the numerical algorithm.   Once this transfer matrix is computed, the RHS of \eqref{FTK} gives an extremely numerically efficient representation of \eqref{CTBA K}.  Even with a $\{\th, \s_1,\s_2\}$ grid containing on the order of $10^4$ points, each iteration of the CTBA can be performed in under 1 second on an ordinary modern computer with only a single core.  With an appropriate damping scheme the method typically converges in around 100 or so iterations as shown in figure \ref{AreaIts}.  
%%%%%%%%%%%%%%%%%%%%%%%%%%%%%%%%%%%
%%%%%%%%%%%%%%%%%%%%%%%%%%%%%%%%%%%
\subsection{String Equations of Motion}\label{Pohl Num}
Let us now turn to an alternative method for computing minimal areas, which is a brute force attack on the string equations of motion, or some reduced variant of them.  This will allow us to perform a final check of the CTBA and the numerical recipe described above.  Since we are only interested in the area (i.e. and not the full string embedding) it is useful to work only with $SO(2,2)$ scalars formed from the $R^{2,2}$ embedding coordinates $X(z, \bar{z})$.  In particular, it is useful to use  work with the variable $\a(z,\bar{z})$ defined by
\beq\label{alpha}
2 e^{2 \a} =  \pd X \cdot \pdb X
\eeq
which appears in the on-shell string action
\beq\label{Afull Pohl}
A_{\text{full}} = 4 \int_{\Sigma} dz d\bar{z} \, e^{2\a}
\eeq
Working in terms of such scalar variables goes by the name of Pohlmeyer reduction \cite{Pohl}, and is actually the starting point for the integrability-based method for computing minimal areas of surfaces with null polygonal boundaries \cite{WilsonLoop}.  After a somewhat lengthy calculation, it follows from \eqref{alpha}, the string equations of motion and Virasoro constraints that $\a$ satisfies the sinh-Gordon equation
\beqy\label{SGE}
\pd \pdb \a = 2 \sinh 2 \a
\eeqy
This equation must be supplemented with boundary conditions that $\a$ approach the straight line solution $\a^{\text{circ}}$ in the vicinity of $z_{\s}$.  The straight line solution satisfies the reduced equation
\beq\label{a circ eqn}
\pd \pdb \a^{\text{circ}} = e^{2 \a^{\text{circ}}}
\eeq
which has a solution for general $z_{\s}$ given by
\beq\label{line soln}
e^{2\a^{\text{circ}}(z,\bar{z})} = \frac{\pd w(z) \pdb \bar{w}(\bar{z})}{(1-w(z) \bar{w}(\bar{z}))^2} 
\eeq
where $w(z)$ is the conformal transformation which maps $z_{\s}$ to the unit circle.  

\begin{figure}[t]
\centering
\includegraphics[width=0.65\linewidth]{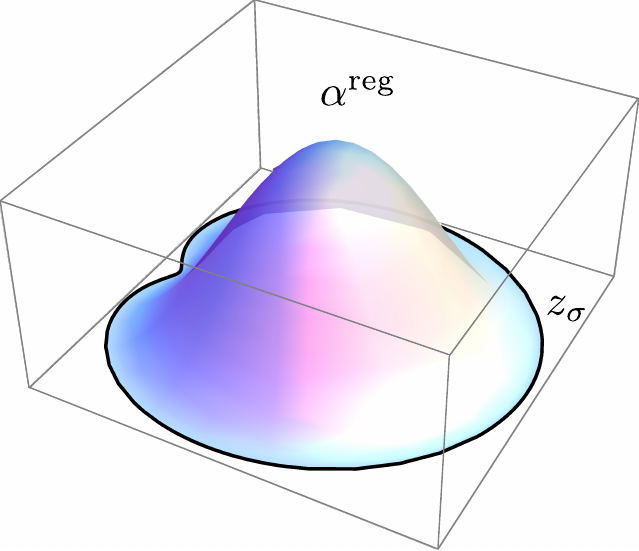}
\caption{Here we plot $\a^{\text{reg}}$ the regulated solution of \eqref{SGE} for the curve $z_{\s}=e^{i \s} + 2/5 \, e^{2 i \s}$ obtained from the relaxation numerics described in this section.  We plot in the domain $\Sigma$ inside of $z_{\s}$. }\label{AlphaRegPlt}
\end{figure} 
Before proceeding to the numerics, we must address some issues of regularization.  First, we must define a suitable function for which to solve.  This is simple: a function which is regular everywhere in $\Sigma$ and on $\pd \Sigma$ is given by
\beq
\a^{\text{reg}} = \a - \a^{\text{circ}}
\eeq
with the boundary conditions $\a^{\text{reg}} \ra 0$ on $\pd \Sigma$.   Second, we must address the regulation of the area \eqref{Afull Pohl} which contains an arc-length divergence due to the warp factor of AdS.  Fortunately, since we know the behavior of the solution near the boundary is given by \eqref{line soln}, it is possible to analytically remove the divergence \cite{ThetaFunctions}.  The trick is to substitute the equations of motion \eqref{SGE} into \eqref{Afull Pohl} and then properly treat a resulting boundary term which contains the divergence.  In the end, the regularized action \eqref{Afull} is given by the beautiful formula \cite{ThetaFunctions}
\beq\label{Martin Area}
A = - 2\pi - 4 \int_{\Sigma} dz d\bar{z} \, e^{-2\a}
\eeq
 
Now we have a well defined numerical problem:  solve \eqref{SGE} written in terms of $\a^{\text{reg}}$ subject to the boundary condition $\a^{\text{reg}} \ra 0$ on $\pd \Sigma$ and then evaluate \eqref{Martin Area}.  For simplicity of implementation, we find it useful to perform the conformal transformation $z \ra w(z)$ which maps the domain $\Sigma$ to the unit disk where it is easy to put down a grid for the numerics.   

Finally, with \eqref{SGE} written in terms of the regular function $\a^{\text{reg}}$ and in the coordinates $(w,\bar{w})$ which live inside the unit disk, we solve this equation using a standard relaxation method which is suitable for the elliptic operator $\pd \pdb$.  One could of course use faster integration schemes based on spectral methods, however we prefer relaxation for its simplicity and stability.  The results of this procedure for an example case are shown in figure \ref{AlphaRegPlt}.  In figure \ref{AreaIts} we compare the results obtained from the numerical integration of the CTBA and the results obtained from integrating the Pohlmeyer reduced equations of motion.  The numbers obtained from the two different methods agree within the expected error of the numerics on both sides. 

As a final comment we note that, as in the case of the CTBA, the Pohlmeyer numerics is completely parameterized in terms of $z_{\s}$.  However, unlike the CTBA, we cannot directly recover the physical boundary curve (i.e. the cross ratios \eqref{Xratios}) after integration.  In order to do that in this approach, one would need to further integrate the equations of motion with \eqref{alpha} to obtain $X$ near the boundary.  Here we are only interested in checking the results of the CTBA and thus this inherent difficulty is of no consequence to us.  Indeed, it clearly demonstrates the advantage of the integrability based approach over direct numerical methods.
\end{appendix}


\begin{thebibliography}{99}
\bibitem{numerics_parameters}
To produce figure \ref{AreaIts} we used a $10 \times 30$ grid for $(\alpha_1,\alpha_2)$.   More points are required in the $\a_2$ direction due to some remaining non-analytic behavior even after the regulation \ref{yreg} (i.e. there is a remaining boundary layer of finite height due to the $+...$ in \eqref{smalls}).  The theta variable is cutoff at $\pm 10$ and contains $32$ grid points.
\bibitem{FootNote_MathieuImp}
In Mathematica we have the following implementation: 
\beqy
M\!c(z) &=& \text{MathieuC}[a(R), R^2, z] \nonumber \\
M\!s (z)&=& \text{MathieuS}[a(R), R^2, z] \nonumber \\
a(R)&=& \text{MathieuCharacteristicA}[1/2, R^2] \nonumber
\eeqy
Recall that $R$ is the radius of the worldsheet curve $z_{\s} = R e^{i \s}$.
\bibitem{ArgBranches}
Here one must take care to properly define the branches of the $\text{arg}$ function appearing in $\varphi_{\ss} = \text{arg}(z_{\ss})$.  This subtlety does not arise in \eqref{CTBA K} due to the $2\pi i$ periodicity of the kernel.
\end{thebibliography}
\end{document}